# (Ultra-)High Energy Muon Neutrino Propagation through the Earth and Induced Muon Energy Distribution near the One Cubic Kilometer Detector

N. Takahashi, Y. Okumura
*Graduate School of Science and Technology, Hirosaki University, Hirosaki 036-8561, Japan*
A. Misaki
*Innovative Research Organization, Saitama University, Saitama 338-8570, Japan*

We calculate high and (ultra-)high energy upward-going muon neutrino propagation through the Earth and the induced muon energy distribution near the one cubic kilometer detector using the Monte Carlo simulation, due to both charged current interaction and neutral one. The initiated neutrino energies on the surface of the Earth are 1PeV, 1EeV and 1ZeV. The mean free paths of (ultra-)high energy neutrino events generated by the deep inelastic scattering may be comparable with the diameter of the Earth or less than it. Therefore, the induced muon production distribution is influenced by the change of the densities interior to the Earth. Furthermore, in such situation, the contribution from the neutral current neutrino interaction to the induced muon production distribution cannot be neglected. We report several examples of the deep inelastic scattered depth of muon neutrino in the Earth and the induced muon energy distribution near the detector.

## 1. INTRODUCTION

It is interesting to study the core and the mantle structure of the Earth by using (ultra-)high energy cosmic ray neutrino beam. For the purpose, we must have the detailed knowledge on the propagation of high energy neutrino through the Earth whose densities are different in position to position. The final aim of this study is to understand the core and mantle structure of the Earth with atmospheric neutrinos.

In the present paper, we give the induced muon energy distribution by such (ultra-)high energy neutrinos near the one cubic kilometer detector.

## 2. IMPORTANCE OF THE NEUTRINO EVENTS DUE TO NEUTRAL CURRENT

When we try to detect neutrinos at (ultra-)high energies where their mean free paths are comparable or even shorter than the diameter of the Earth, we must take into consideration two important factors, namely, the change of the density interior to the Earth and the contribution from neutral current interaction, because the former is directly related to the magnitude of the neutrino cross sections and the latter is related to non-negligible to the final charged current events from a successive neutral current interaction.

In the present paper, we adopt the preliminary Earth Model [1] as for the density inside the Earth and differential cross section of the neutrinos at (ultra-)high energies obtained by Gandhi et al [2].

In (ultra-)high energies, neutrinos can choose either of following two reactions:

$$\nu_\mu + N \rightarrow \mu + X \qquad (1)$$

$$\nu_\mu + N \rightarrow \nu_\mu + X \qquad (2)$$

First of all, it should be noticed that we detect neutrinos only through the charged particles, namely, muons or electrons and the produced charged particles can be detected only near the detector. Even if (ultra-) high energy charged particles can be produced far from the detector, such charged particles cannot reach the detector. For the moment, we are interested in muons, because the effective volume for muon detection is far bigger than that for electrons.

We are now interested in the upward neutrinos. Due to the shortening of the mean free paths of the neutrinos as the increase of their energies, generally speaking, frequency of the neutral current events may be not negligible than that of the charged current events. Because the charged current interactions at (ultra-)high energies produce the muons at the location far from the detector and such muons never reach the detector, while the neutral current interactions at (ultra-)high energies may produce neutrinos whose energies are smaller than that of primary neutrino and the second generation neutrino has longer mean free path compared with that of the first generation as the result of the decrease of energy. Thus, the second generation neutrino may produce the third generation neutrino with longer mean free path via neutral current interaction and so on. In conclusion, the neutral current interaction may have not negligible probability to produce muons near the detector in some case, compared with the corresponding charged current interaction. Of course, at the final stage of the interaction, neutrinos produced from the last generation of the neutral current interaction must take part in the charged current interactions which produce finally muons. Otherwise, the original neutral current interactions are never detected.

In the present paper, we give the frequencies of neutral current events which produce finally the muon via charged current interaction at their last stage together with frequencies of the charged current events.

## 3. METHOD FOR CALCULATION

As we must examine the change of the density inside the Earth, we adopt the differential method in the sense of the Monte Carlo methods. We obtain frequency distribution of both charged current events and neutral current events for the incident neutrinos on the surface of the Earth with monochromatic primary energies.

### 3.1. The charged current interaction

**[a]** We start from an (ultra-)high energy neutrino which is expected to produce muon somewhere in the Earth via charged current interaction. The probability which occur within in $dx(\rho)$ is given by $dx(\rho)/\lambda_{CC}(E_\nu)$, where $\lambda_{CC}(E_\nu)$ denotes a mean free path of the neutrino with the energy $E_\nu$ due to charged current interaction. We sample $\xi$, a uniform random number *(0, 1)*. If $\xi$ is smaller than $dx(\rho)/\lambda_{CC}(E_\nu)$, then, we judge the interaction concerned occurs. Next, we sample $\xi$, a uniform random number *(0, 1)* and determine $E_\mu$, the energy of the muon produced due to charged current interaction, utilizing the following formula.

$$\xi = \frac{\int_{E_{\min}}^{E_\mu} D_{CC}(E_\nu, E_\mu)dE_\mu}{\int_{E_{\min}}^{E_{\max}} D_{CC}(E_\nu, E_\mu)dE_\mu} \qquad (3)$$

, where $D_{CC}(E_\nu, E_\mu)dE_\mu$ denotes the charged current differential cross section for deep inelastic scattering.

Thus, the interaction point of the neutrino and the energy of the produced muon are recorded. After that, we sample another neutrino with the same energy and repeat the same procedure until we attain the enough statistics. If $\xi$ larger than $dx(\rho)/\lambda_{CC}(E_\nu)$, then, we judge that the interaction concerned does not occur and we try the same procedure in the previous $dx(\rho)$ to new and next step $dx(\rho)$ and so on.

### 3.2. The neutral current interaction

**[b]** We start from the neutrino with a given energy whose the mean free path is given by $\lambda_{NC}(E_\nu)$. In the first $dx(\rho)$, we compare $\xi$, a sampled uniform random number with $dx(\rho)/\lambda_{NC}(E_\nu)$. If $\xi$ is smaller than $dx(\rho)/\lambda_{NC}(E_\nu)$, then, we judge the neutral current interaction occurs within $dx(\rho)$. Next, we determine the produced neutrino energy due to neutral current interaction, utilizing the following equation:

$$\xi = \frac{\int_{E_{\min}}^{E_{\nu'}} D_{NC}(E_\nu, E_{\nu'})dE_{\nu'}}{\int_{E_{\min}}^{E_{\max}} D_{NC}(E_\nu, E_{\nu'})dE_{\nu'}} \qquad (4)$$

, where $D_{NC}(E_\nu, E_{\nu'})dE_{\nu'}$ denotes the differential cross section for neutral current.

**[c]** The neutrino thus produced has whose energy is $E_{\nu'}$ two choices, namely, one is the charged current interaction and the other is the neutral current interaction. Here we introduce the following quantities:

$$\xi_i = \frac{\sigma_{NC}(E_{\nu'})}{\sigma_{CC}(E_{\nu'}) + \sigma_{NC}(E_{\nu'})} \qquad (5)$$

Now, we sample new uniform random number $\xi$. If $\xi < \xi_i$, then, we judge the neutral current interaction occurs and we go to the procedure [b] with new energy $E_{\nu'}$. Otherwise, we judge the charged current interaction occurs and we go to the procedure[a].

## 4. THE INTERACTION POINT DISTRIBUTION OF THE NEUTRINO EVENTS

From the experimental point of view, the charged particles (muons) due to neutrino interactions which are produced near the detector can be measurable. Therefore, it is important to understand induced muon energy distribution near the detector.

Supposing that the incident neutrinos penetrate through the Earth from its surface, we pursue the neutrino concerned in the stochastic manner by the Monte Carlo method which turns to interact with in *(t, dt)*. The density of the Earth as the function of the depth traversed by the neutrinos for different zenith angles is given in Figure 1. In Figure 2 and Figure 3, we give the mean free paths due to charged current as the functions of the distance from the detector for $\cos\theta = -1.0$ (vertically upward) and $\cos\theta = -0.7$, respectively.

We calculate the interaction point distribution of the neutrino events in two cases, namely, 1) due to charged current (CC) only and 2) due to charged current plus neutral current [CC+NC]. However, it should be noticed that [CC+NC] is the natural choice compared with the choice of CC. The incident neutrinos start from the surface opposite side of the observation point and passing through the Earth toward the detector which is located at 1500 meters in water.

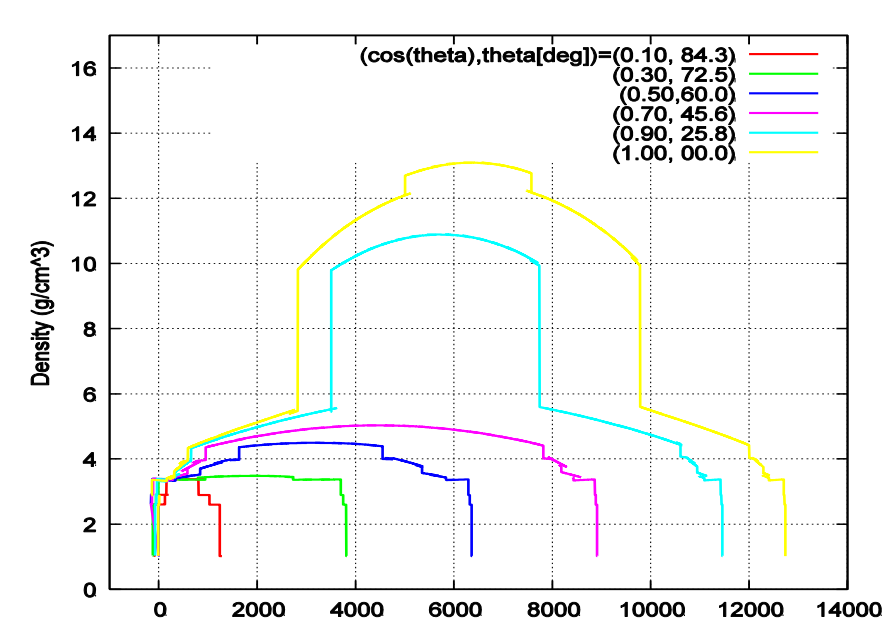

Figure 1: Density Profile of the Earth calculated from the preliminary Earth Model ([1-2]).

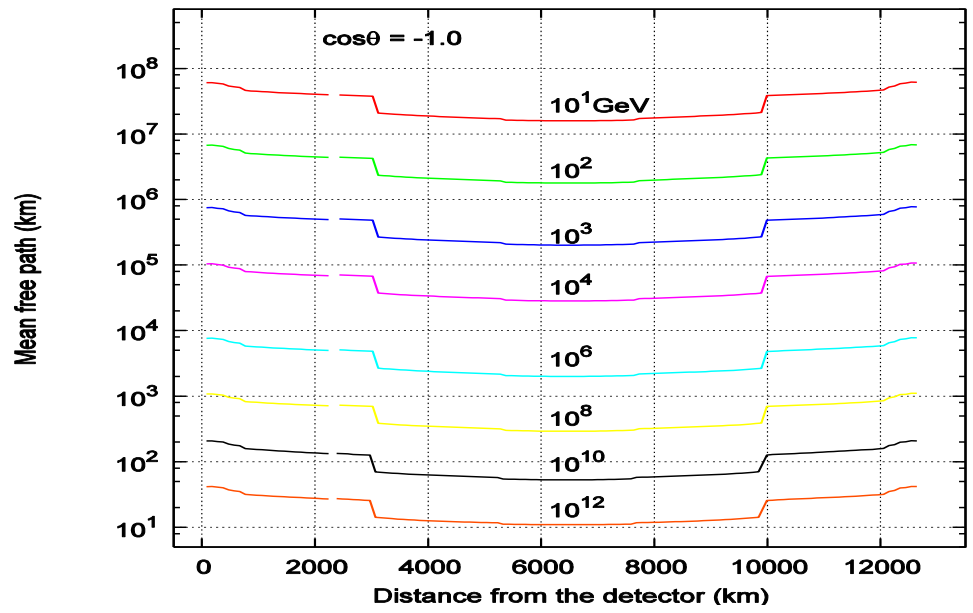

Figure 2: Neutrino mean free paths as the function of the depths for $\cos\theta = -1.0$.

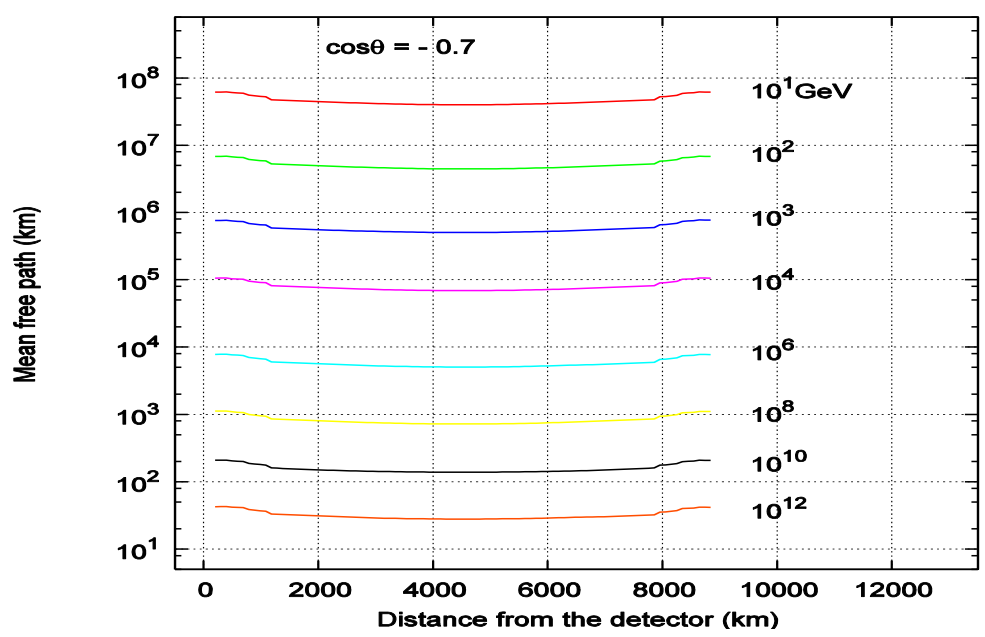

Figure 3: Neutrino mean free paths as the function of the depths for $\cos\theta = -0.7$.

## 4.1. The neutrino events due to charged current

Here, we assume that only charged current interaction exsists. Let us suppose that neutrinos interact with the matter through charged current. The probability for neutrinos to interact in *(t, t+dt)* is given as,

$$P_{CC,\mathrm{int}}(E_\nu, t, \cos\theta) = \left(1 - \frac{dt}{\lambda_{CC,1}(E_\nu, t_1, \rho_1)}\right) \times \cdots$$

$$\cdots \times \left(1 - \frac{dt}{\lambda_{CC,n-1}(E_\nu, t_{n-1}, \rho_{n-1})}\right) \times \left(\frac{dt}{\lambda_{CC,n}(E_\nu, t_n, \rho_n)}\right) \quad (6)$$

Here, $\lambda_{CC,n}$ denotes the mean free path of the neutrino event due to charged current, whose energy and density are $E_\nu$, and $\rho_n$ at the depth *t*, respectively.

The same probability can be calculated by the Monte Carlo method. We utilize the both methods for obtaining the interaction point for the neutrinos due to the charged current.

## 4.2. The neutrino events due to neutral current in addition to those due to charged current

The neutrino events due to neutral current appear as the product if the last stage is realized via the charged current interaction. In other words, neutrino events due to neutral current are defined as the events in which the first interaction is due to the neutral current and the last interaction is due to the charged current.

In Figure 4, we show the interaction point distribution for the neutrino with 10TeV due to charged current for $\cos\theta = -1.0$ by Eq.(1) (blue point). The same distribution is obtained by the Monte Carlo method (green point). The agreement between them is quite well, which guarantees the validity of our Monte Carlo method.

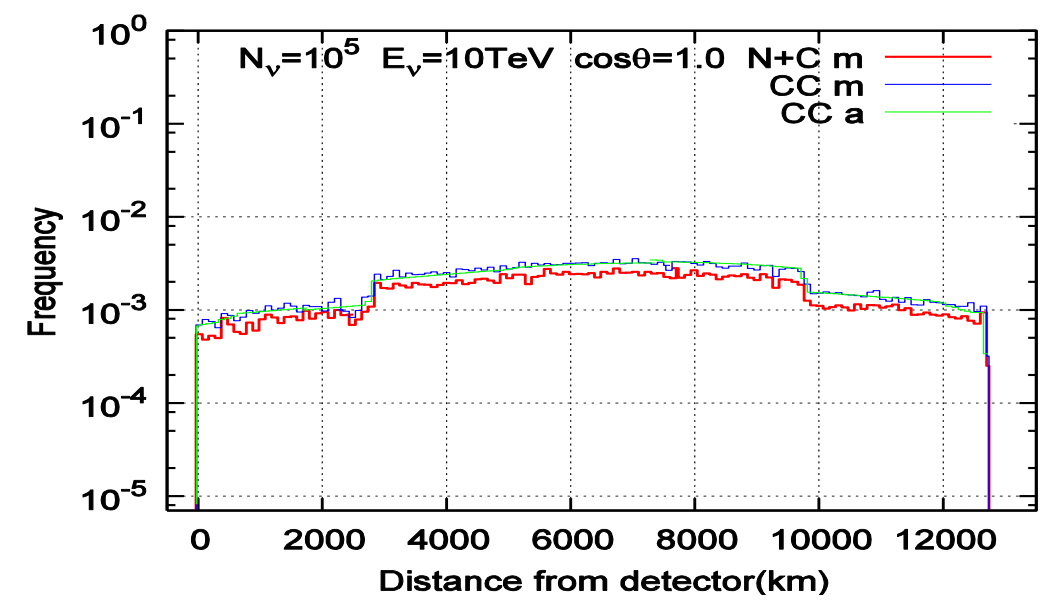

Figure 4 Interaction point distribution for CC and [CC+NC] with 10TeV neutrino.

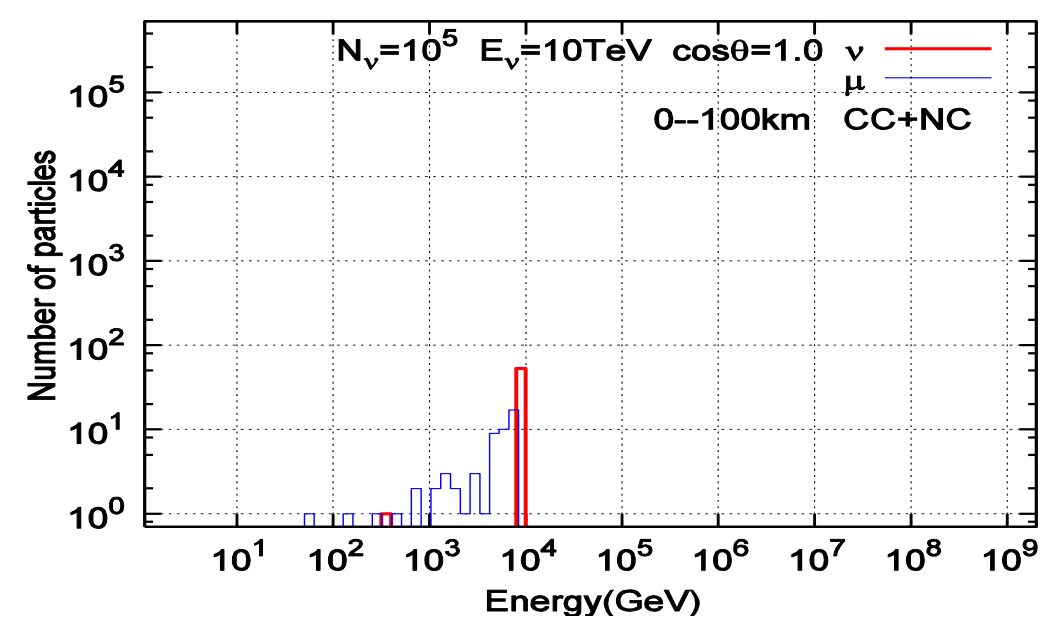

Figure5: The energy spectra for neutrinos and the produced muons initiated by 10TeV neutrino within 100km from the detector.

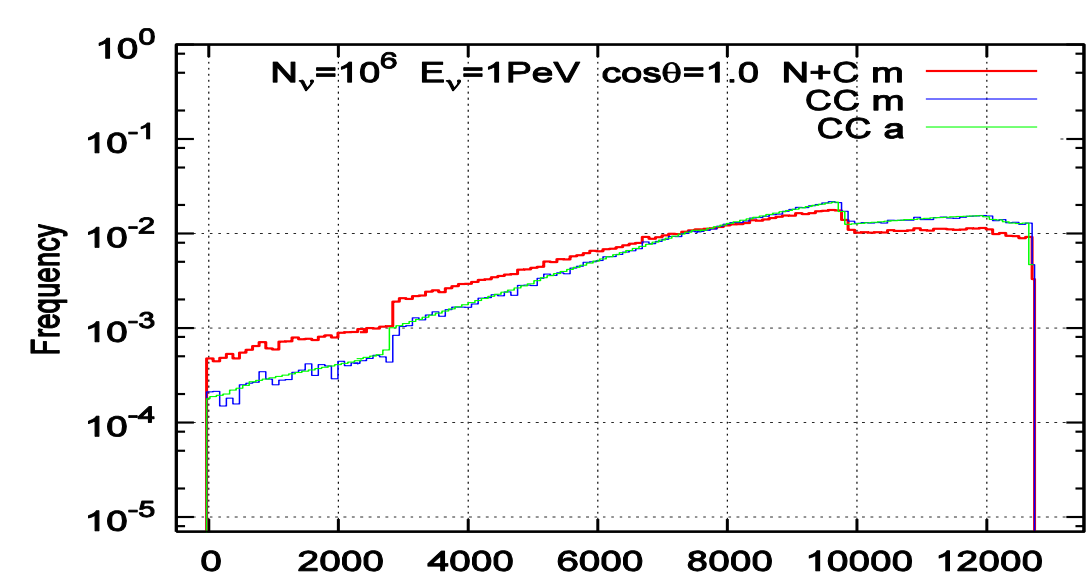

Figure 6: Interaction point distributions for CC and [CC+NC] with 1PeV neutrino.

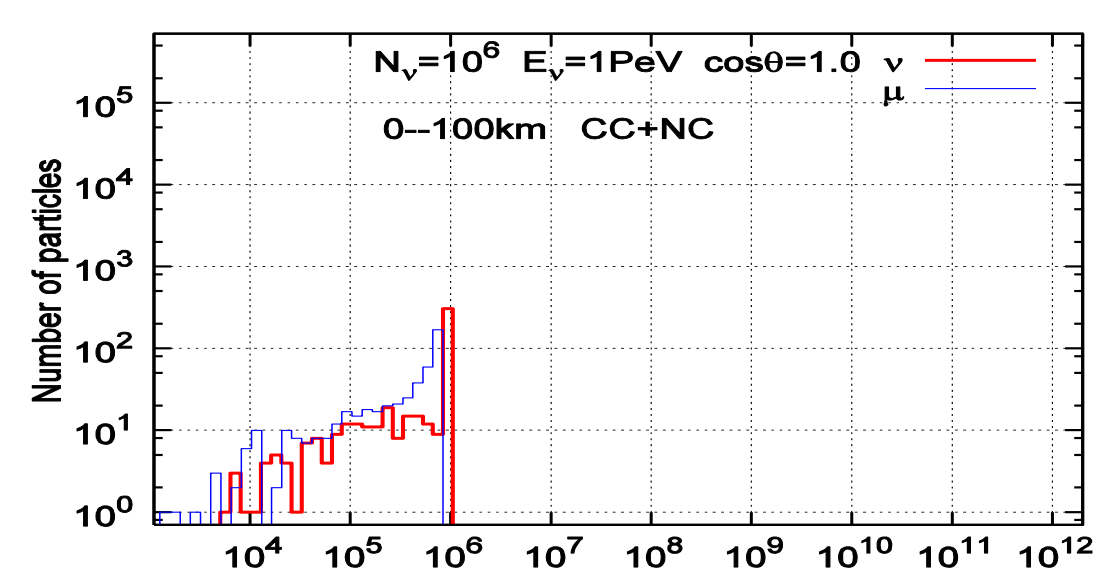

Figure 7: The energy spectra for the neutrinos and the produced muon initiated by 1PeV neutrino within 100 km from the detector.

Furthermore, we add the corresponding distribution for neutrino due to [CC+NC] by the Monte Carlo method in Figure 4.

It is clear from the comparison of the red points (charged current plus neutral current) with blue(green) points (charged current) that neutrino events via neutral current cannot be neglected compared with the contribution from the neutrino events due to neutral current.

There are two characteristics in Figure 4. The first one is that there are two abrupt changes at both ~10000km and ~3000km which correspond to the entrance of the core of the Earth and the exit of the core. The second one is that the interaction point distributions are approximately flat, which denote the mean free path of 10TeV neutrino is larger than the diameter of the Earth (roughly by four times). In Figure 5, we show the energy spectra for neutrinos and produced muons which exist within 100 km from the detector) for [CC+NC].

In Figure 6, we give the similar result with the initiated energy of 1PeV for $\cos\theta = -1.0$ to Figure 4. The abrupt changes in the interaction point distribution occur at the same distances as in Figure 5. In Figure 7, we give the similar results to Figure 5.

In Figure 8, we give the similar result with the initiated energy of 1EeV to Figure 6. Also, Figure 9 corresponds to Figure 7. It is clear from the Figure 8 as for comparison of red points with blue (green) points that all neutrino events which are produced within 100 km from the detector are produced via neutral currents. In Figure 9, we give the energy spectra for neutrinos and the produced muons due to neutral current only, because the corresponding spectra due to charged current disappear due to shorter mean free paths.

In Figure 10, we give the interaction point distribution with same initiated energy in Figure 8, but for $\cos\theta = -0.2$. In Figure 11, we give the energy spectra for neutrinos and the produced muons within 100 km from the detector which corresponds to Figure10.

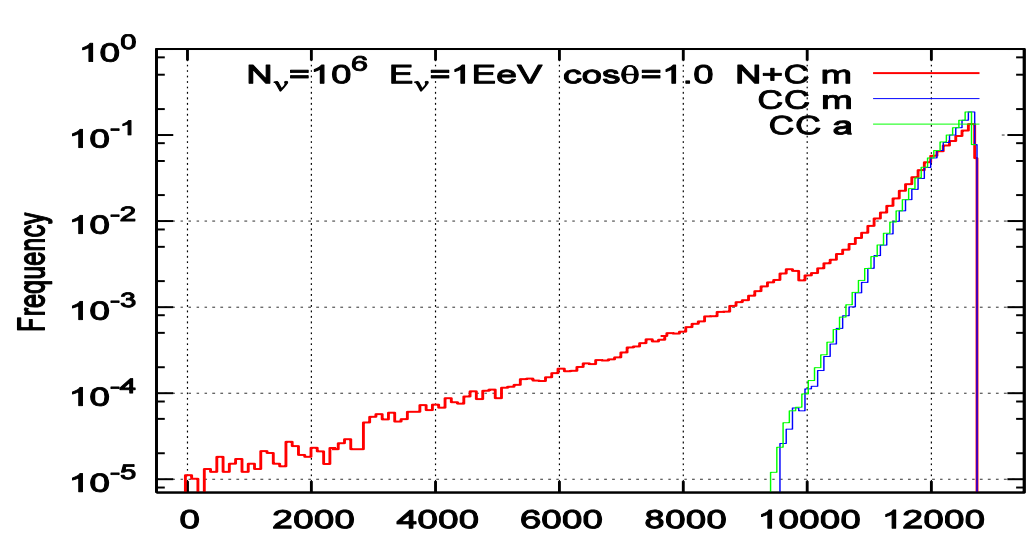

Figure 8: Interaction point distribution for CC and [CC+NC] with 1EeV neutrino.

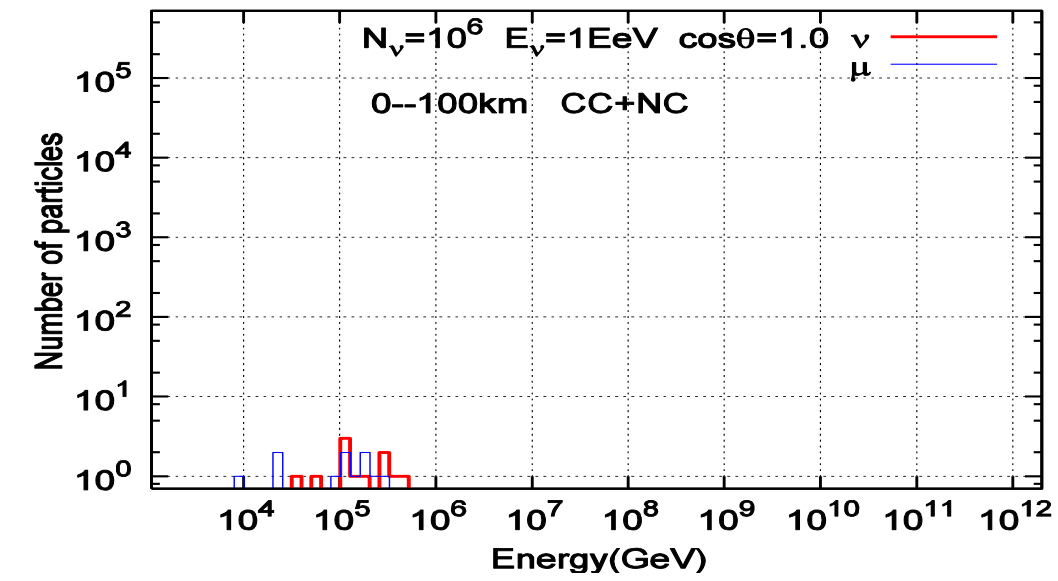

Figure 9: The energy spectra for the neutrino and the produced muon initiated by 1EeV neutrino within 100 km from the detector.

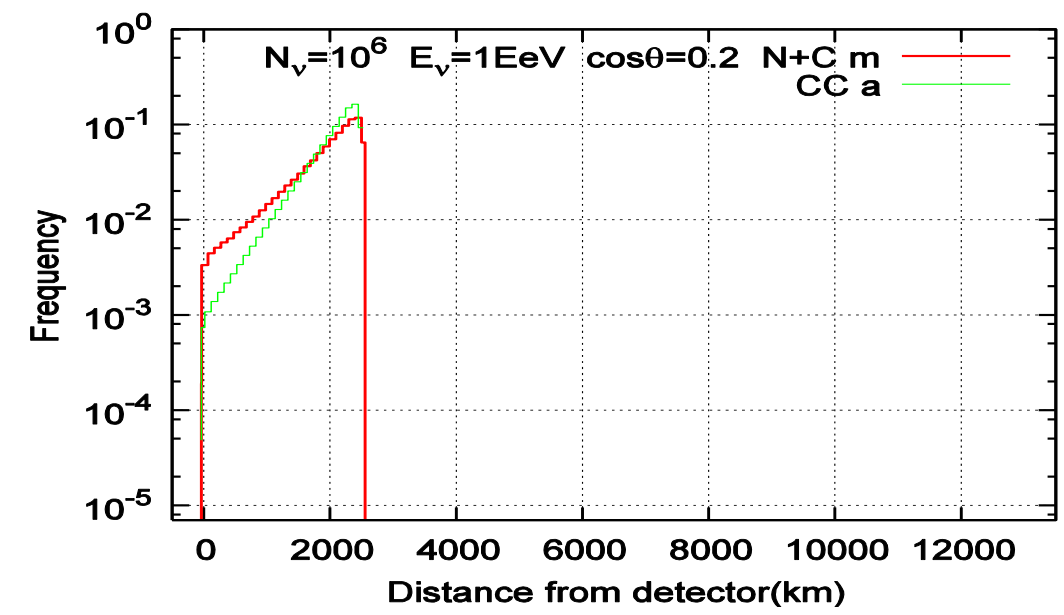

Figure 10: Interaction point distributions for CC and [CC+NC] with 1EeV neutrino.

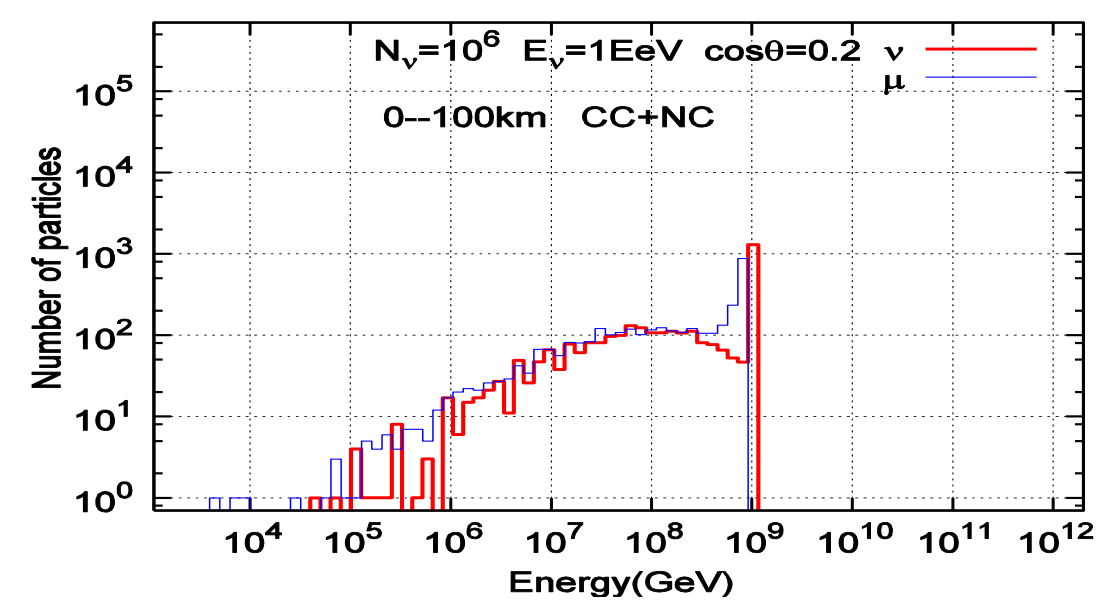

Figure 11: The energy spectra for the neutrino and the produced muon initiated by 1EeV neutrino within 100 km from the detector.

## 5. SUMMARY

It is clear from Figure 4, 6, 8 and 10 that the contribution of the interaction point distribution due to neutral current cannot be neglected, on the contrary to L'abbate et al [3].